\documentstyle[12pt]{article}
\begin{document}
\title{Comment on "Why quantum mechanics cannot be formulated as a
Markov process"}
\author{Piotr Garbaczewski  and Robert Olkiewicz\\
Institute of Theoretical Physics, University of Wroc{\l}aw,\\
PL-50 204 Wroc{\l}aw, Poland}
\maketitle
\hspace*{1cm}
\begin{abstract}
In the paper with the above title,  D. T. Gillespie [Phys. Rev. A 49,
1607, (1994)] claims that the theory of Markov stochastic processes
cannot provide an adequate mathematical  framework for quantum
mechanics. In conjunction with the specific quantum dynamics
considered there, we give a general analysis of the associated
dichotomic jump processes.\\
If we assume that Gillespie's "measurement probabilities" \it
are \rm the transition probabilities of a stochastic process, then the
process must have an invariant (time independent) probability measure.
Alternatively,  if we demand  the probability measure of the process
to follow the quantally implemented (via the Born statistical
postulate) evolution, then we  arrive at the  jump process
which \it can \rm be interpreted as a Markov
process if restricted to a suitable duration time.
However, there is no corresponding  Markov process consistent with
the  $Z_2$ event space assumption, if we require  its  existence
for all times $t\in R_+$.
\end{abstract}

\vskip0.2cm

Before, \cite{olk}, we have contested the general statement due to
Gillespie \cite{gil} about the generic contradiction between the
probabilistic concepts appropriate for quantum theory and those proper
to the common-sense theory of stochastic (in particular, Markov)
processes.
Our argument was based on   invoking the standard,  configuration space,
Schr\"{o}dinger picture quantum dynamics which, if combined with
the Born statistical interpretation postulate, allows for a consistent
description  in terms of Markov processes of diffusion type. In
conformity with the rich theory developed so far,
\cite{nel,combe,zambr,blanch,olk1,olk2,olk3}.

However, our arguments did not pertain to stochastic jump processes
which were the main objective of  Ref. \cite{gil}.
Let us therefore consider  a simple
two-level quantum system  undergoing the
Schr\"{o}dinger evolution:
$${\psi (t)=exp(-ict)\: cos\: \omega t\: |1> - i\: exp(-ict)\: sin\:
\omega t\: |2>}\eqno (1)$$
and  concentrate on its  probabilistic analysis, with an additional
motivation coming from the series of papers due to other authors
\cite{falco,ang,ang1,ang2,dje,dje1}, where the Markov property has
been attributed to analogous dynamical problems, see however
Ref.\cite{garb}.
We shall slightly simplify (1) by rescaling the dimensional
constants to achieve $\omega =1$.

The discussion of Ref. \cite{gil}  departs from the following
\it epistemological \rm  input: "If the
system is known to be in state $1$ at time $s$, then the
probabilities of finding the system at any time $t>s$ to be in
states $1$ and $2$ are $cos^2(t-s)$ and $sin^2(t-s)$ respectively,
and similarly that if the system is known to be in state $2$
at time $s$ then  the probabilities of finding the system at any time
$t>s$ to be in states $2$ and $1$ are $cos^2(t-s)$ and $sin^2(t-s)$
respectively". These "measurement probabilities" are then utilised as
\it transition probabilities  \rm of a certain (presumed to be
consistently defined) stochastic jump process, with the outcome
that the quantum
mechanical evolution is at variance with the canonical form of the
master  equation appropriate for the problem (Eqs. (13) in Ref.
\cite{gil}).

The above transition  probabilities constitute a two by two
transition matrix $p^G(t,s)$ with elements:
$${p^G_{11}(t,s)=cos^2(t-s)=p^G_{22}(t,s)}\eqno (2)$$
$$p^G_{12}(t,s)=sin^2(t-s)=p^G_{21}(t,s)$$

To avoid any possible confusion, let us recall (see e.g. Ref.
\cite{gih}) that a stochastic
process on
$Z_2$, if considered on  a finite time interval, say $[0,T]$,
is to be given by a hierarchy of transition probabilities
(they are an
easy transcription of those conventionally utilized in the framework
of continuous processes):
$${p(\sigma ,t)\; ,\; 0\leq t\leq T}\eqno (3)$$
$$p(\sigma _1,t_1|\sigma _2,t_2)\; ,\; 0\leq t_2<t_1\leq T$$
$$p(\sigma _1,t_1|\sigma _2,t_2,\sigma _3,t_3)\; , \; 0\leq
t_3<t_2<t_1\leq T$$
and  so on, where each index $\sigma $ equals either $1$ or $2$.

In the above, $p(\sigma ,t)=\mu \{ \omega \in \Omega: X_t(\omega
)=\sigma \}, \: \sigma =1,2$ defines a probability measure of the
stochastic process on $Z_2$ i.e.  probabilities with which the
dichotomic random variable takes  values
along  conrete sample (jumping) paths in the event space $\Omega $.
The probability measure of the process is then propagated
(or left invariant) by the
transition probability $p(\sigma _1,t_1|\sigma _2,t_2)$.

 The  probabilities  (3) have to satisfy  the so called
 consistency conditions:\\

(i) $\sum _{\sigma } p(\sigma ,t)=1$ \\

(ii) $ \sum _{\sigma _1} p(\sigma _1,t_1|\sigma _2,t_2) =1 $
and  $ \sum _{\sigma _2} p(\sigma _1,t_1|\sigma _2,t_2)
p(\sigma _2,t_2) =
p(\sigma _1,t_1)$    \\

(iii)  $\sum _{\sigma _1}
p(\sigma _1,t_1|\sigma _2,t_2,\sigma _3,t_3)
=1$\\

$\sum _{\sigma _2} p(\sigma _1,t_1|\sigma _2,t_2,\sigma
_3,t_3)p(\sigma _2,t_2|\sigma _3,t_3) =
p(\sigma _1,t_1|\sigma _3,t_3)$, \\

$\sum _{\sigma _3} p(\sigma _1,t_1|\sigma _2,t_2,\sigma
_3,t_3) p(\sigma _2,t_2|\sigma _3,t_3) p(\sigma _3,t_3) =
p(\sigma
_1,t_1|\sigma _2,t_2) p(\sigma _2,t_2)$ \\

etc.

For a Markov process we would have $p(\sigma _1,t_1|\sigma
_2,t_2,\sigma _3,t_3)=p(\sigma _1,t_1|\sigma _2,t_2)$ in which case
(iii) would reduce to a single identity
$${\sum _{\sigma _2} p(\sigma _1,t_1|\sigma _2,t_2) p(\sigma
_2,t_2|\sigma _3,t_3) =p(\sigma _1,t_1|\sigma _3,t_3)}\eqno (4)$$
i.e. the Chapman-Kolmogorov equation. Then, the hierarchy is closed
and the process is completely specified by giving its initial
probability measure \it and \rm its transition probabilities.

Let us point out that in Ref. \cite{gil} the  random dynamics was
characterised exclusively in terms of transition probabilities  and
 with no reference to a probability measure of the process.
The probabilistic description of random jumps on $Z_2$
patterned after  \cite{falco,ang}  is given in
terms of the "probability vector" (probability measure in the
present case):
$$ p(t)={\left( \begin{array}{c} p_1(t)\\
p_2(t)\end{array} \right)}\eqno (5)$$
$$p_1(t)\geq 0\; ,\; p_2(t)\geq 0\; ,\; p_1(t)+p_2(t)=1$$
and the transition probability
$${p(t_1,t_2)=\left( \begin{array}{cc}
p_{11}(t_1,t_2) & p_{12}(t_1,t_2)\\
p_{21}(t_1,t_2) & p_{22}(t_1,t_2)
\end{array} \right)}\eqno (6)$$
$$p_{ij}(t_1,t_2)=p(i,t_1|j,t_2)\; ,\; i,j=1,2$$
$$p_{ij}(t_1,t_2)\geq 0\; , \; \sum_{i} p_{ij}(t_1,t_2)=1$$
$$\sum_{j} p_{ij}(t_1,t_2) p_j(t_2)=p_i(t_1)$$
The last identity is equivalent to
$p(t_1,t_2)p(t_2)=p(t_1)$,
understood as the matrix-vector operation in the linear space.

If Gillespie's transition matrix $p^G(t,s)$, (2),  is to define
a consistent
stochastic process, in view of its  breaking the Chapman-Kolmogorov
identity,  the higher rank conditional probabilities need
to be introduced.
Unfortunately, they are not given in Ref. \cite{gil}.

Nevertheless, let us take for granted that this supplementary step
 can be made so that  $p^G(t,s)$ \it is \rm a transition matrix
 of a well defined (nonMarkovian) stochastic process
 $X_t$  with values in $Z_2$.
It is natural to ask for the probability measure $\mu $ of this process
i. e. for its probability vector $p(t)$.
It must  satisfy the consistency (in fact, propagation) condition
$${p^G(t,s)p(s)=p(t)}\eqno (7)$$
for all $s<t$.
Let us analyze the issue in some detail.

If $p(0)$ is an arbitrary initial density, we can always write
$$ p(0)={\left( \begin{array}{c} a\\
1-a \end{array} \right)}\eqno (8)$$
with $a\in [0,1]$.
Then, for all $s>0$ we have:
$${p(s)=p^G(s,0)p(0)={\left( \begin{array}{c}
a cos^2s + (1-a)  sin^2s\\
a sin^2s + (1-a) cos^2s \end{array} \right)}}\eqno (9)$$
and  for every $t>s$
$${p(t)=p^G(t,0)p(0)={\left( \begin{array}{c}
a cos^2t + (1-a)  sin^2t\\
a sin^2t + (1-a) cos^2t \end{array} \right)}}\eqno (10)$$
On the other hand, there holds:
$${p(t)=p^G(t,s)p(s)=}\eqno (11)$$
$${\left( \begin{array}{c}
cos^2(t-s) [a cos^2s + (1-a) sin^2s] +  sin^2(t-s) [a sin^2 s +
(1-a) cos^2s]\\
sin^2(t-s) [a cos^2s + (1-a) sin^2s]  +
cos^2(t-s) [a sin^2s + (1-a) cos^2s]\end{array} \right)}$$
Since (10) and (11) must be equal, we get the identity:
$${(a-{1\over 2})sin2(t-s) sin2s = 0}\eqno (12)$$
to be valid for all $0<s<t$. It implies  $a={1\over 2}$ and
consequently
$${p(0)={\left( \begin{array}{c}
1/2\\
1/2 \end{array} \right)}}\eqno (13)$$
is the only admissible initial choice of $p(0)$.

Moreover, $p^G(t,s)$ is a symmetric matrix, which implies the following
important property:
if for some $t_0$ we
deal with  $p(\sigma ,t_0)= const$ which is independent of
$\sigma =1,2$, then
for all $t>t_0$ there  is $p(\sigma ,t)=p(\sigma ,t_0)=const$.
This follows from the observation:
$${p(\sigma ,t)=\sum _{\sigma _0} p(\sigma ,t|\sigma _0,t_0) p(
\sigma
_0,t_0)=
const\; \sum _{\sigma _0} p(\sigma ,t|\sigma _0,t_0)=}\eqno (14)$$
$$const \; \sum _{\sigma _0} p(\sigma _0,t|\sigma ,t_0) = const $$

As a consequence, by utilizing $p^G(t,s)$ as a transition probability
appropriate for the stochastic process, we would arrive at the
process, whose  probability measure is conserved in time
$${p(0)=p(t)=  {\left( \begin{array}{c}
1/2\\
1/2 \end{array} \right)}}\eqno (15)$$
for all $t>0$.

The random dynamics induced by $p^G(t,s)$ is thus
appropriate exclusively  for systems with an invariant
probability measure.
It  certainly has nothing in common with the
quantally implemented evolution of the probability measure associated
(via the Born postulate)  with the explicit solution (1),
 i. e. with the probability vector $p(t)$ whose components read:
 $p(1,t)=cos^2t,\: p(2,t)=sin^2(t)$.

We believe, that  a consistent approach towards a  probabilistic
reinterpretation of the quantum dynamics proper should result
in the construction of a  stochastic process
(Markovian if possible) which is compatible with the
quantum Schr\"{o}dinger picture evolution (1).
This issue has received
attention in the literature, see e.g. Refs. 3-9  and Refs. 10-16  in
particular.

Let us follow this, alternative with respect to the reasoning
of Ref. \cite{gil}, idea  and demonstrate  that
the restriction to symmetric transition matrices (Gillespie's case,
see e.g. Ref. \cite{carlen} for more detailed discussion)
   implies that there is no consistent  Markov jump process
 which can be associated with the Schr\"{o}dinger dynamics (1) for
 all times.

Indeed, for the time dependent probability vector $p(t)$,
 at time $t_0={\pi \over 4}$ we have
$${p(t_0)={\left( \begin{array}{c} cos^2(\pi /4)\\
sin^2(\pi /4) \end{array} \right) }= {\left( \begin{array}{c} 1/2 \\
1/2 \end{array} \right) }}\eqno (16)$$
and the previous result follows for times
exceeding $\pi /4$.

However,  we can construct a
 Markov process running in the finite  time interval $[0,\pi /4]$.

We have $p(t)={\left( \begin{array}{c} cos^2t \\
sin^2t \end{array} \right) }$. Let us define $p(t,s)$:
$${p_{11}(t,s)=p_{22}(t,s)={{cos^2t-sin^2s}\over {cos\: 2s}}  }\eqno (17)$$
$$p_{12}(t,s)=p_{21}(t,s)=
1-p_{11}(t,s)={{cos^2s - cos^2t}\over {cos\: 2s}}$$
where $0\leq s<t \leq {\pi \over 4}$.
All these   matrix coefficients are nonnegative in the time interval
 $[0,{\pi \over 4}]$.
Moreover, for $0\leq s<t<u\leq {\pi \over 4}$, by inspection
(with some help of trigonometric identities) we find  the
Chapman-Kolmogorov equation to be valid
$${p(u,t)p(t,s)=p(u,s)}\eqno (18)$$
and so the Markov property is established.
Obviously, the propagation formula
$${p(t,s){\left( \begin{array}{c} cos^2s \\
sin^2s \end{array} \right) }= {\left(  \begin{array}{c} cos^2t \\
sin^2t \end{array} \right) }}\eqno (19)$$
holds true. Also, we have
$lim_{t\downarrow s}p(t,s)= I$ where $I$ denotes the unit
two-by-two matrix. It corresponds to  the matrix element property
$p(i,s|j,s)=\delta _{ij}$ for all $i,j=1,2$.

The major steps of our analysis, (3)-(19), did not rely on
any "quantum measurement" epistemology and merely  invoked
mathematical features of stochastic jump processes on $Z_2$.
However, the main difference
between the approach of Ref. \cite{gil}  and this of Refs.
\cite{falco,ang} is rooted in the preferred choice of the connection
between mathematics and physics (here, pertaining to the concept of
measurement in quantum theory).

The  theory of stochastic processes is normally regarded by
physicists  as a macroscopic theory in the sense that one can
probe the system without significantly perturbing it. Then,
it is not surprising that such a theory may be  viewed 0as
inconsistent with the ordinary quantum mechanics. This line
of thought is followed in Ref. \cite{gil}.

On the other hand, instead of viewing the stochastic process as
a description of one system evolving from $t= -\infty$ to
$t=+\infty $, one might view  it as a theory for an infinite
ensemble of systems, all starting either in the same initial state
or with some initial probability distribution over the possible
states at time $t=0$. The ensemble is allowed to evolve until we
possibly decide to stop the systems. Viewed this way, the theory of
stochastic processes might very well describe the evolution of a
quantum system between measurements (compare e.g. in this
connection the two time localization framework of Ref.
\cite{schulman}). And that is the admissible
interpretation of our previous discussion. In this case, on
suitable time scales we have found the Markov property to persist.

Let us stress that if the zero point of the time axis is taken to
be the time of the measurement on the system, where the
system \it was \rm found to be in state $1$, then there is no
disagreement between  our example (17)-(19) and the results of Ref.
\cite{gil}.  The propagation from $s=0$ can be safely extended
to an arbitrary time instant $T>0$ interpreted as a subsequent
measurement of the system. However, to analyze the time
evolution of the probability measure $p(t)$ between those  fixed
time interval boundaries, we must define the transition probability
$p(t,s)$ for \it all \rm  intermediate time instants. Obviously,
this corresponds to transitions between \it unobserved \rm
intermediate states and, as shows (18), for not too long time
intervals of interest such  stochastic interpolation  is Markovian
until interrupted (terminated) by the measurement.

\vskip0.2cm

{\bf Acknowledgement}: Both authors receive a financial support from 
the  KBN research grant No 2 P302 057 07. We would like to express
our gratitude to the anonymous Referee for interesting comments.

\end{document}